\documentclass[sn-mathphys]{sn-jnl}

\usepackage{pdfpages}
\graphicspath{{./figs/}{./png/}}

\jyear{2022}%
\theoremstyle{thmstyleone}%
\theoremstyle{thmstyletwo}%

\theoremstyle{thmstylethree}%

\newcommand{\brac}[1]{\langle #1 \rangle}
\newcommand{\Fig}[1]{Fig.~\ref{#1}}

\newcommand{\bra}[1]{\langle #1\rangle}

{}
{}
{}

{}
{}
{}
{}
{}
{}
{}
{}
{}
{}
{}
{}
{}
{}
{}
{}
{}
{}

{}
{}
{}

{}
{}
{}
{}

{}

{}
{}

\newcommand{\uu}{\mbox{\boldmath $u$} {}}

\newcommand{\BB}{\mbox{\boldmath $B$} {}}

\newcommand{\JJ}{\mbox{\boldmath $J$} {}}
\newcommand{\SSS}{\mbox{\boldmath $S$} {}}
\newcommand{\AAA}{\mbox{\boldmath $A$} {}}

\newcommand{\ff}{\mbox{\boldmath $f$} {}}

\newcommand{\nab}{\mbox{\boldmath $\nabla$} {}}

\newcommand{\SSSS}{\mbox{\boldmath ${\sf S}$} {}}

\newcommand{\DD}{{\rm D} {}}

\def\Ma{\mbox{\rm Ma}}

\def\Pm{\mbox{\rm Pr}_{\rm M}}
\def\Rm{\mbox{\rm Re}_{\rm M}}
\def\Rmc{\mbox{\rm Re}_{\rm M}^{\rm crit}}
\def\Rey{\mbox{\rm Re}}

\def\cs{c_{\rm s}}

\def\kf{k_{\rm f}}
\def\kB{k_{\rm b}}
\def\kBs{k_{\rm bs}}
\def\kn{k_{\nu}}
\def\ke{k_{\eta}}
\def\kM{k_{\rm M}}

\def\Brms{B_{\rm rms}}

\def\urms{u_{\rm rms}}

\begin{document}

\title[Small-scale dynamos at low Prandtl numbers]{Numerical evidence
  for a small-scale dynamo approaching solar magnetic Prandtl numbers}


\author*[1]{\fnm{J\"orn} \sur{Warnecke}}\email{warnecke@mps.mpg.de}

\author[2,1,3]{\fnm{Maarit J.} \sur{Korpi-Lagg}}

\author[2,4]{\fnm{Frederick A.} \sur{Gent}}

\author[2]{\fnm{Matthias} \sur{Rheinhardt}}

\affil*[1]{\orgname{Max-Planck-Institut f\"ur Sonnensystemforschung}, \orgaddress{\street{Justus-von-Liebig-Weg 3}, \city{G\"ottingen}, \postcode{D-37077}, \country{Germany}}}

\affil[2]{\orgdiv{Department of Computer Science}, \orgname{Aalto University}, \orgaddress{\street{PO Box 15400}, \city{Espoo}, \postcode{FI-00\ 076}, \country{Finland}}}

\affil[3]{\orgdiv{Nordita}, \orgname{KTH Royal Institute of Technology \& Stockholm University}, \orgaddress{\street{Hannes Alfv\'ens v\"ag 12}, \city{Stockholm}, \postcode{SE-11419}, \country{Sweden}}}

\affil[4]{\orgdiv{School of Mathematics, Statistics and Physics}, \orgname{Newcastle University}, \orgaddress{ \city{Newcastle upon Tyne}, \postcode{NE1 7RU}, \country{UK}}}

\abstract{%
Magnetic fields on small scales are ubiquitous in the universe.
Though they can often be observed in detail,
their generation mechanisms are not fully understood. 
One possibility is the so-called small-scale dynamo (SSD).
Prevailing numerical evidence, however, appears to indicate that
an SSD is unlikely to exist at very low magnetic Prandtl 
numbers ($\Pm$) such as are present in the Sun and other cool stars.
We have performed high-resolution
simulations of isothermal forced turbulence
employing the lowest $\Pm$ values so far achieved.
Contrary to earlier findings,  the SSD turns out to be not only
 possible for $\Pm$ down to 0.0031,
but even becomes increasingly easier to excite
for $\Pm$ below $\simeq\,$0.05.
We relate this behaviour to the known hydrodynamic
phenomenon referred to as the bottleneck effect.
Extrapolating our results to solar values of
$\Pm$ indicates that an SSD would be possible under such conditions.
}

\maketitle

Astrophysical flows are considered
to be susceptible to two types of dynamo instabilities.
Firstly, a large-scale dynamo (LSD) is excited 
by flows exhibiting helicity, or more generally, lacking mirror-symmetry,
due to rotation, shear, and/or stratification. It generates coherent, dynamically significant, 
magnetic fields on the global scales of the object in question \cite{BS05}.
Characteristics of LSDs vary depending on the dominating generative effects, 
such as differential rotation in the case of the Sun.
Convective turbulence provides
both generative and dissipative effects \citep{Ch20},
and their presence and astrophysical relevance is no longer strongly
debated. 

The presence of the other type of dynamo instability, namely the small-scale 
or fluctuation dynamo (SSD), however, remains controversial in
solar and stellar physics. 
In an SSD-active system, the magnetic field is generated
at scales comparable
to, or smaller than the characteristic scales of the turbulent flow,
enabled by chaotic stretching of field lines at high magnetic Reynolds number \cite{Child+Gilbert95}.
In contrast to the LSD,  excitation of an SSD requires significantly
stronger turbulence \cite{BS05}.
Furthermore, it has been theorized that it becomes more difficult to excite SSD at very low
magnetic Prandtl number
$\Pm$
\cite{BC04,Hau04,SCMM04,SOCMPY07,ISCM07,Schober+12,BHLS18}, 
the ratio of kinematic viscosity $\nu$ and
magnetic diffusivity $\eta$.
In the Sun, $\Pm$ can reach values as low as $10^{-6}$--$10^{-4}$
\cite{Stix02},
thus seriously repudiating whether an SSD can at all be present.
Numerical models of SSD in near-surface solar convection typically
operate at $\Pm\approx 1$ \cite{C99,VS07,KKMW15,HRY15s,Rempel14,R18,RS22}
and thus circumvent the issue of low-$\Pm$ dynamos.

A powerful SSD may potentially have a large impact on the dynamical
processes in the Sun. It can, for example, influence the angular momentum
transport and therefore the generation of differential rotation
\cite{KKOWB17,HK21},
interact with the LSD \cite{TC13,BSB16,SB16,HRY16,VPKRSK21}
or contribute to coronal heating via enhanced photospheric
Poynting flux \cite{Rempel17}. 
Hence, it is of great importance to clarify 
whether or not an SSD can exist in the Sun.
Observationally, it is still debated whether
the small-scale magnetic field on the surface
of the Sun has contributions from the SSD or is 
solely due to the tangling of the large-scale magnetic field by the turbulent motions
\cite{KBSB10,BLS13,LCM14,BRLOSD19,FR21,KLOT22}.
However, these studies show a slight preference of 
the small-scale fields to be cycle-independent.  
SSD at small $\Pm$ are also important for the interiors 
of planets and for liquid metal experiments \cite{Tobias21}.

Various numerical studies have reported increasing difficulties in
exciting the SSD when decreasing $\Pm$
\cite{SCMM04,SHBCMM05,BHLS18}, confirming the theoretical predictions.   
However, current numerical models
reach only $\Pm=0.03$ using explicit physical diffusion 
or slightly lower 
(estimated) $\Pm$, relying on artificial hyperdiffusion \cite{SOCMPY07,ISCM07}.
To achieve even lower $\Pm$, one needs to increase the grid
resolution massively, see also \cite{tobias_cattaneo_boldyrev_2012}.
Exciting the SSD requires a magnetic Reynolds number
($\Rm$) typically larger than
100; hence, e.g., $\Pm=0.01$ implies 
a fluid Reynolds number
$\Rey=10^4$, where
$\Rey=\urms\ell/\nu$, $\urms$ being the
volume integrated root-mean-squared velocity, 
$\ell$ a characteristic scale of the velocity,
and $\Rm=\Pm\Rey$.
In this paper, we take this path and lower $\Pm$
significantly using 
high-resolution simulations.

\begin{figure*}[t!]
\begin{center}
\includegraphics[width=\columnwidth]{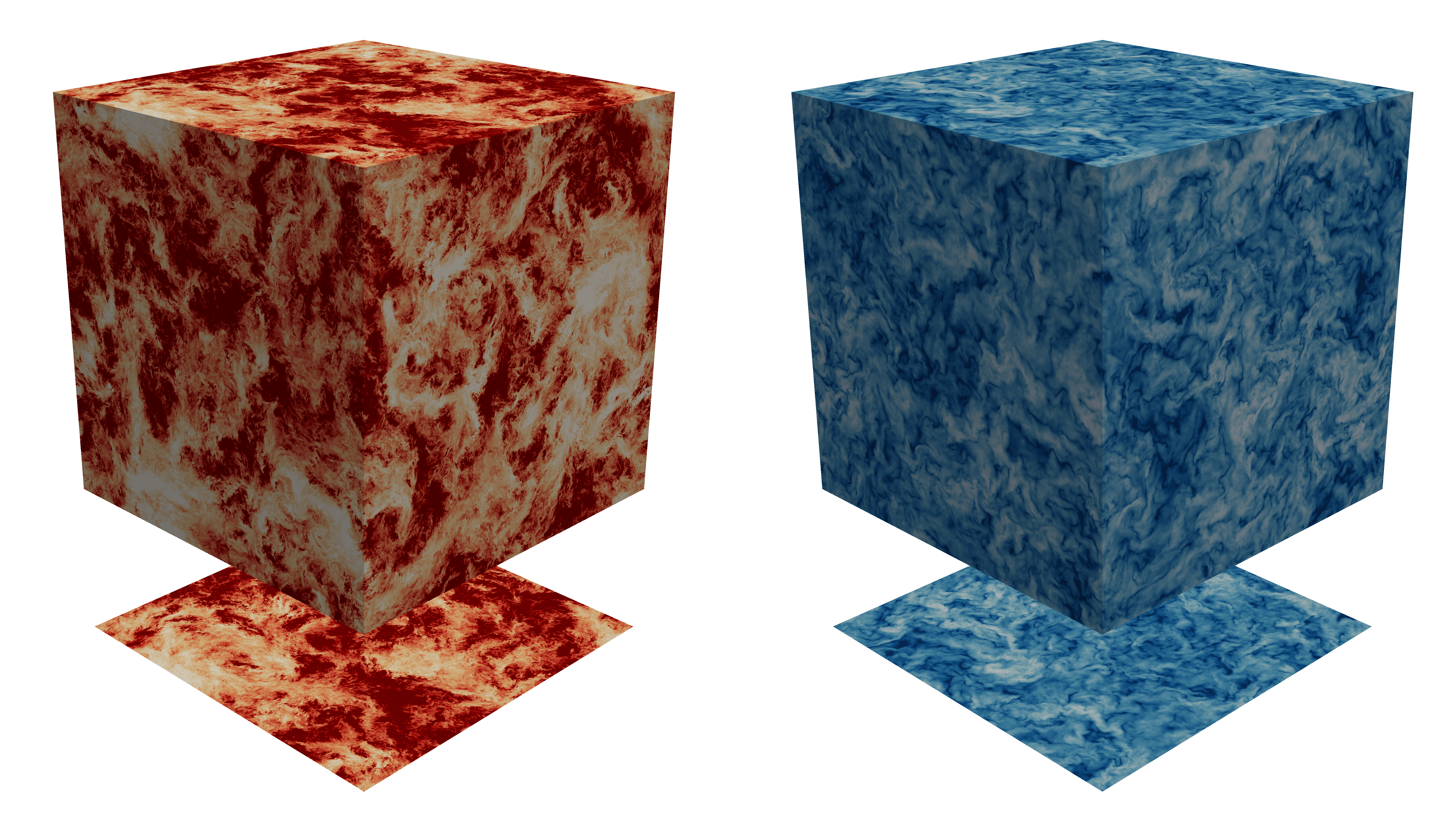}
\end{center}\caption[]{
        {\bf Visualisation of flow and SSD solution.}
        Flow speed (left) and magnetic field strength (right) from a high resolution
	SSD-active run with $\Rey=18200$ and $\Pm = 0.01$ on the surface of the simulation box.
}\label{fancy}
\end{figure*}

\section*{Results}
We include simulations with resolutions of 256$^3$ to 4608$^3$ grid points
and $\Rey=46$ to $33000$.
This allows us to explore the parameter space
from $\Pm=1$ to $0.0025$, which is closer to the solar value than has
been investigated in previous studies.
For each run, we measure the growth rate $\lambda$ of the 
magnetic field in its
kinematic stage and determine whether or not an SSD is being excited. 

To afford an in-depth exploration of the effect of $\Pm$,
we omit large-scale effects such as stratification, rotation and
shear.
We avoid the excessive integration times, required to simulate convection,
by driving the turbulent flow explicitly under
isothermal conditions.
Our simulation setup consists of a fully periodic box 
with a random volume force, see Online Methods
for details;
the flow exhibits a Mach number of around 0.08.
In Fig.~\ref{fancy}, we visualize the velocity and  magnetic
fields of one of the highest resolution and
Reynolds number cases.
As might be anticipated for low $\Pm$ turbulence, the flow exhibits much finer,
fractal-like structures than the magnetic field.
Note, that all our results refer to the kinematic stage of the SSD, where the
magnetic field strength is far too weak to influence the flow and otherwise arbitrary.

\begin{figure*}[t!]
\begin{center}
\includegraphics[width=\columnwidth]{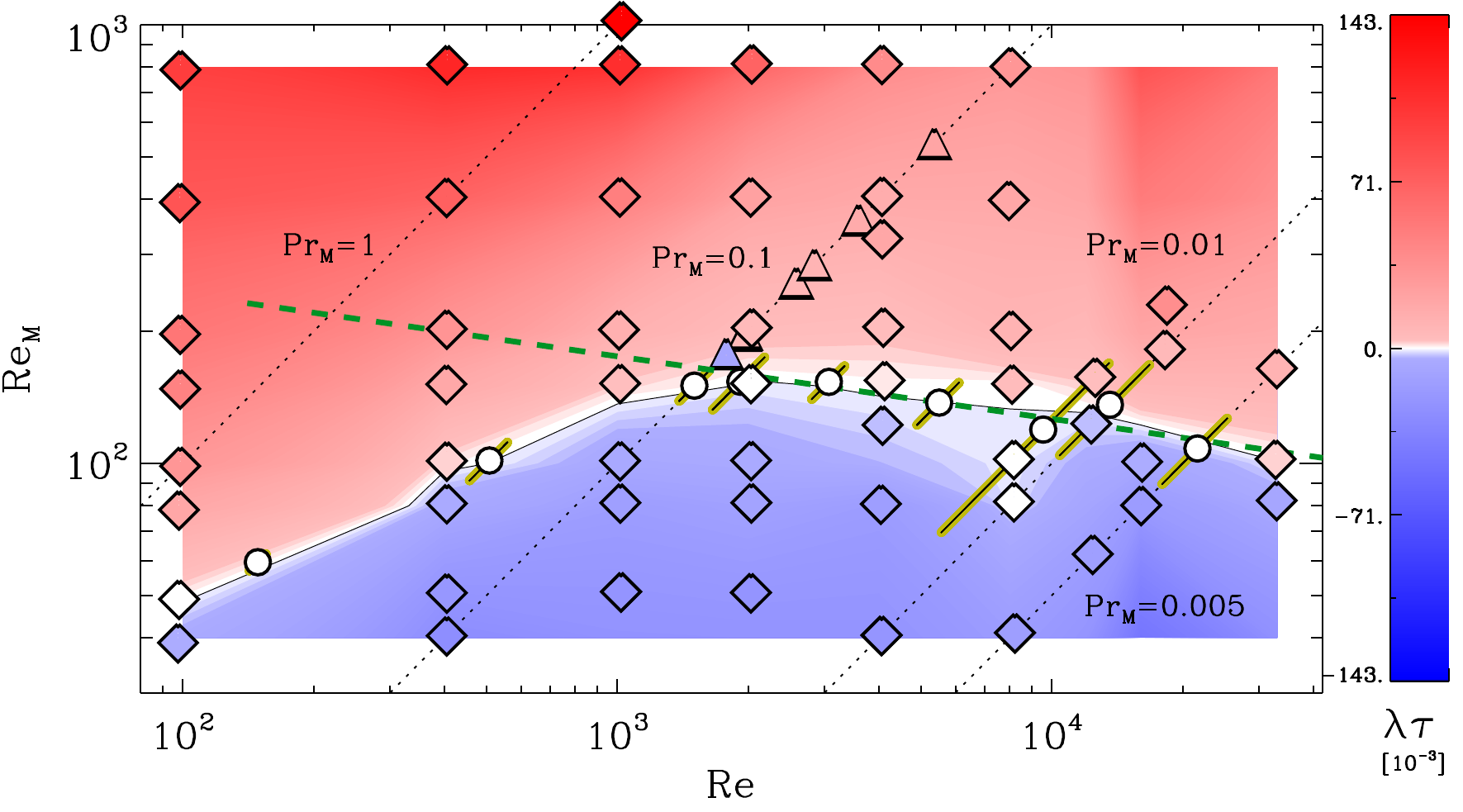}
\end{center}\caption[]{
{\bf Small-scale dynamo
growth rate as function of the fluid and
magnetic Reynolds numbers ($\Rey$ and $\Rm$).}
The diamonds represent the results
of this work and the triangles those of \cite{BHLS18}.
The color coding indicates the value of the 
normalized growth rate
 $\lambda\tau$ with $\tau=1/\urms\kf$,
a rough estimate for the turnover time.
Dotted lines indicate constant magnetic Prandtl number $\Pm$.
White circles indicate zero growth rate for certain $\Pm$,
obtained from fitting for the critical magnetic Reynolds number, as shown in \Fig{lambda2};
fitting errors are signified by yellow-black bars.
see Supplementary Material, Section~5.
The background colors including the thin black line (zero growth) are
assigned via linear interpolation of the simulation data.
The green dashed line shows the power-law fit of the critical $\Rm$ for $\Pm\le0.08$,
with power 0.125, see \Fig{lambda2}b.
}\label{lambda}
\end{figure*}

\subsection*{Growth rates and critical magnetic Reynolds numbers}
In Fig.~\ref{lambda} we visualize
the growth rate $\lambda$ as function of 
$\Rey$ and $\Rm$.
We find positive growth rates for all sets of runs with
constant $\Pm$ if $\Rm$ is large enough. 
$\lambda$ increases always with increasing $\Rm$ as expected.
Surprisingly, the growth rates are significantly lower   
within the interval from $\Rey=2000$ to 10000 
than below and above.
With the $\Rm$ values used, this maps roughly to a $\Pm$ interval from about 0.1 to 0.04. 
 
The growth rates for $\Pm=0.1$ match very well the ones from
\cite{BHLS18}, indicated by triangles.
From Fig.~\ref{lambda}, we clearly see that
the critical magnetic Reynolds number $\Rm^{\rm crit}$,  defined by growth rate $\lambda=0$,
first rises as a function of $\Rey$ and then 
falls for $\Rey>3\times10^3$, see the thin black line.
Looking at $\Rm^{\rm crit}$ as a function of magnetic Prandtl number $\Pm$, 
it first increases with decreasing $\Pm$
and then decreases for $\Pm<0.05$.
Hence, an SSD is easier to excite here than for $0.05<\Pm<0.1$.
We could even find a nearly marginal, positive growth rate for
$\Pm=0.003125$.
The decrease of $\lambda$ at low $\Pm$ 
is an important result as the SSD
was believed to be even harder \cite{BC04,Schober+12} or at least equally
hard \cite{SOCMPY07,ISCM07} to excite when $\Pm$ was decreased further from previously
investigated values.
The growth rates 
agree qualitatively with the earlier work at low
$\Pm$ \cite{SCMM04,SOCMPY07,ISCM07}.

\begin{figure*}[t!]
\begin{center}
\includegraphics[width=\columnwidth]{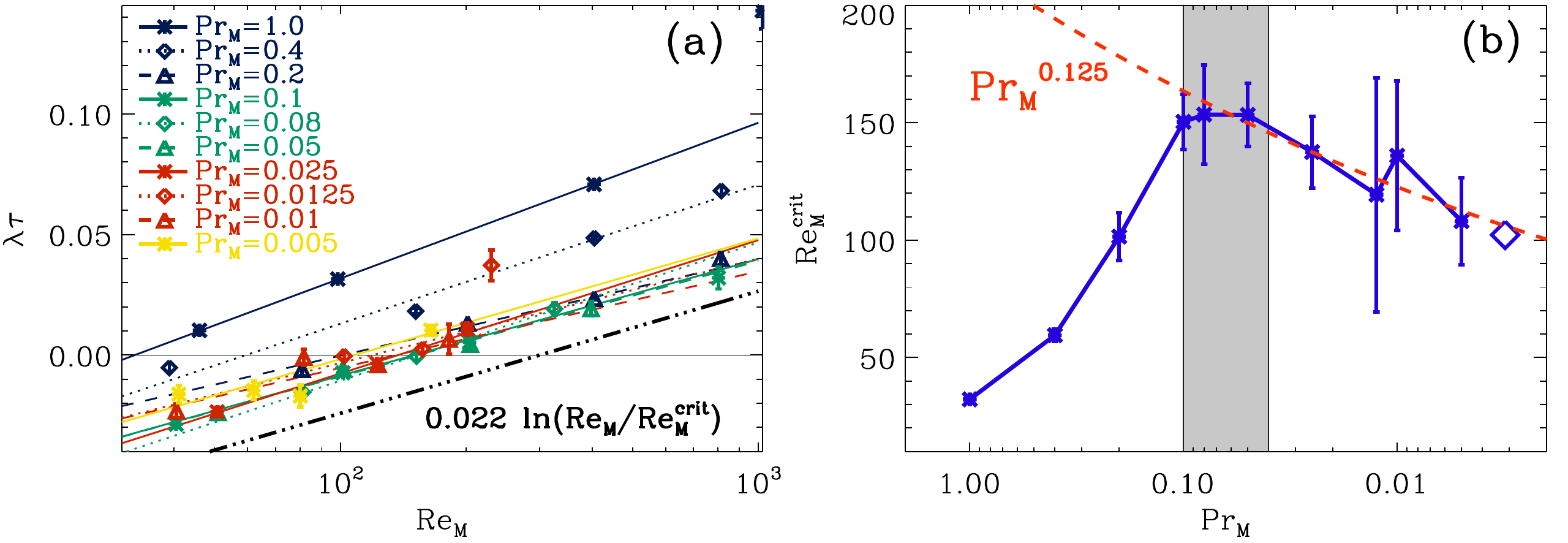}
\end{center}\caption[]{
{\bf Growth rate and critical Reynolds number.}
Panel (a):
normalized growth rate $\lambda\tau$ as function of magnetic Reynolds number $\Rm$ for
simulation sets with fixed magnetic Prandtl number $\Pm$,  
indicated by different colors. 
Logarithmic functions $\lambda\tau \propto      
\ln{(\Rm/\Rm^{\rm crit})}$ according to
\cite{KR97,KR12} were fitted separately to the individual sets,
as indicated by the colored lines;
see the dashed-dotted line for the mean slope.
Panel (b): critical magnetic Reynolds number $\Rm^{\rm crit}$ as function of $\Pm$ 
obtained from the fits in panel (a).
Error bars show the fitting error,
see Supplementary Material, Section~5.
The diamond indicates a run with growth rate $\lambda\approx 0$,
hence its $\Rm$ represents $\approx\Rm^{\rm crit}$ for the used $\Pm=0.003125$.
The red dashed line is a power law fit
$\Rm^{\rm crit}\propto \Pm^{0.125}$,
valid for $\Pm\lesssim0.08$.
The grey shaded area indicates the $\Pm$ interval
where the dynamo is hardest to excite ($\Rm^{\rm crit}\gtrsim150$).
}\label{lambda2}
\end{figure*}

For a more accurate determination of $\Rm^{\rm crit}$,
we next plot the growth rates 
for fixed $\Pm$ as a function of $\Rm$, see
Fig.~\ref{lambda2}(a).
The data are consistent with
$\lambda\propto\ln{(\Rm/\Rm^{\rm crit})}$
as theoretically predicted
\cite{KR97,KR12}.
Fitting accordingly,
we are able to determine $\Rm^{\rm crit}$ as a function of $\Pm$, see
Fig.~\ref{lambda2}(b).
The latter plot clearly shows that there are three distinct
regions of dynamo excitation:
When $\Pm$ decreases in the range $1\ge\Pm\ge0.1$ it becomes much harder to
excite the SSD.  In the range $0.1\ge\Pm\ge0.04$, 
excitation is most difficult with little variation of $\Rm^{\rm crit}$. 
For $\Pm\le0.04$, it again becomes easier as $\Pm$ reduces.
In \cite{SOCMPY07,ISCM07}, the authors already found an
indication of $\Rm^{\rm crit}$ to level-off with decreasing $\Pm$,
however only when using artificial hyperdiffusion.
Similarly, with our error bars, a constant $\Rm^{\rm crit}$
cannot be excluded for $0.01<\Pm<0.1$.
However, at
$\Pm=0.005$, the error bar allows to conclude that $\Rm^{\rm crit}$ is here lower than at $\Pm=0.05$.
This again confirms our result that $\Rm^{\rm crit}$ is decreasing
with $\Pm$ for very low $\Pm$.

For $\Pm\le0.05$, 
the decrease of $\Rm^{\rm crit}$ with $\Pm$ 
can be well represented by the power-law 
$\Rm^{\rm crit}\propto\Pm^{0.125}$.
Extrapolating this to the Sun and solar-like stars would lead to
$\Rm^{\rm crit}\approx40$ at $\Pm=10^{-6}$, 
which means that we could expect an SSD to be present.
For increasing $\Rey$, by decreasing $\nu$, it would be reasonable to assert
that
the statistical properties of the flow and hence $\Rmc$ become independent of $\Pm$.
However, episodes of non-monotonic behavior of $\Rm^{\rm crit}$ 
when approaching this limit
cannot be ruled out.

The well-determined 
$\Rm^{\rm crit}$ dependency on $\Pm$
together with its error
bars and the power-law fit have been added to Fig.~\ref{lambda},
and agree very well with the thin black line for $\lambda=0$
interpolated from the growth rates.

\begin{figure}[t!]
\begin{center}
\includegraphics[width=\columnwidth]{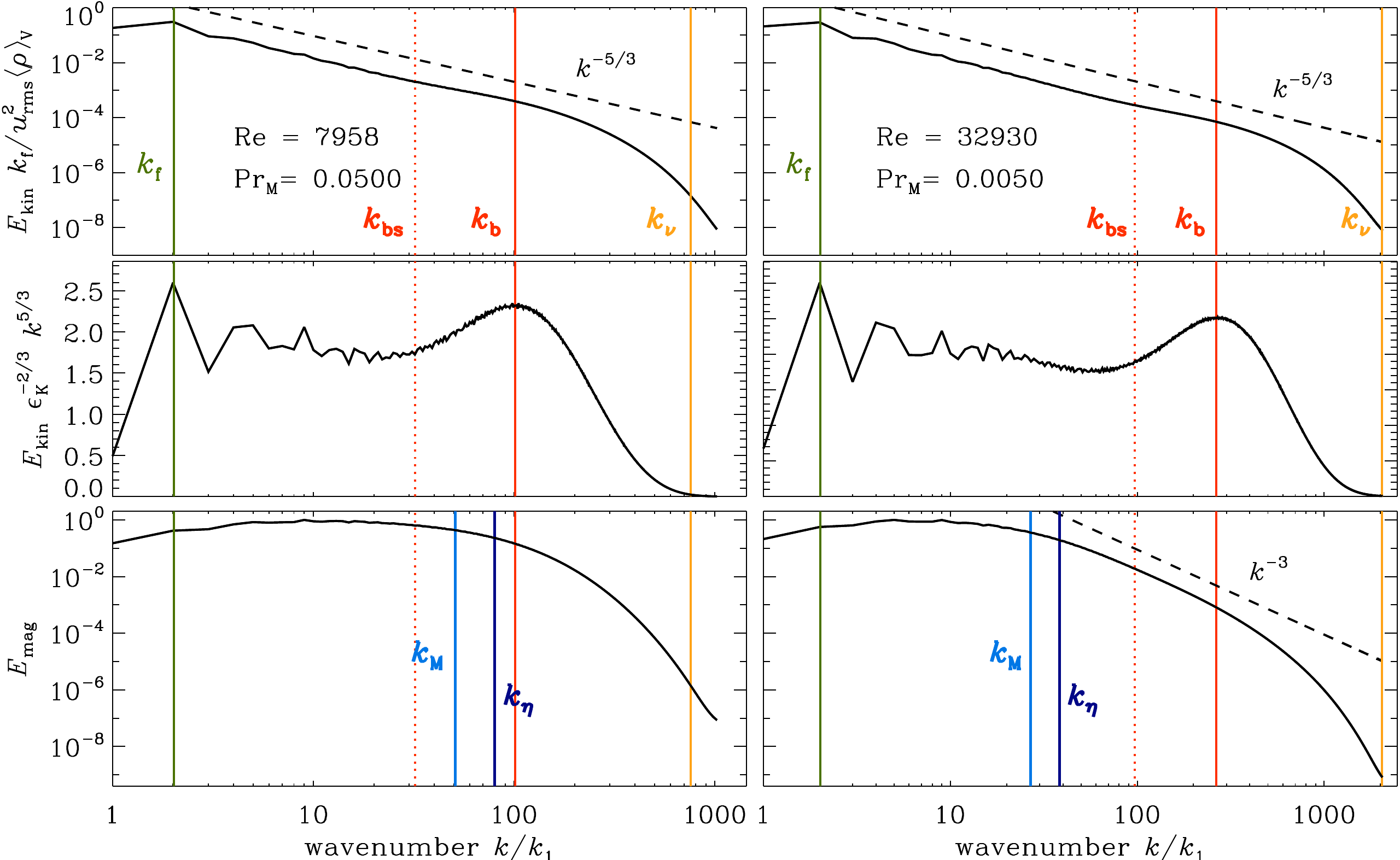}
\end{center}\caption[]{
{\bf Energy spectra.}
Kinetic (top row) and magnetic (bottom row) energy spectra for two exemplary runs with
$\Rey=7958$, $\Pm=0.05$ (left column) and $\Rey=32930$, $\Pm=0.005$ (right column).
In the middle row, the kinetic spectra are compensated by $k^{5/3}$.
Vertical lines indicate the forcing wavenumber $\kf$ (green solid),
the wavenumber of the bottleneck's peak 
$\kB$ (red solid)  and its starting point $\kBs$ 
(red dotted), the viscous dissipation wavenumber 
$\kn$ (orange), the ohmic dissipation wavenumber 
$\ke=\kn\Pm^{3/4}$ (dark blue) and the 
characteristic magnetic wavenumber $\kM$
(light blue).
All spectra are averaged over the kinematic phase
whereupon each individual magnetic spectrum was normalized by its maximum, thus
taking out the exponential growth.
}\label{spec}
\end{figure}

\begin{figure}[t!]
\begin{center}
\includegraphics[width=0.75\columnwidth]{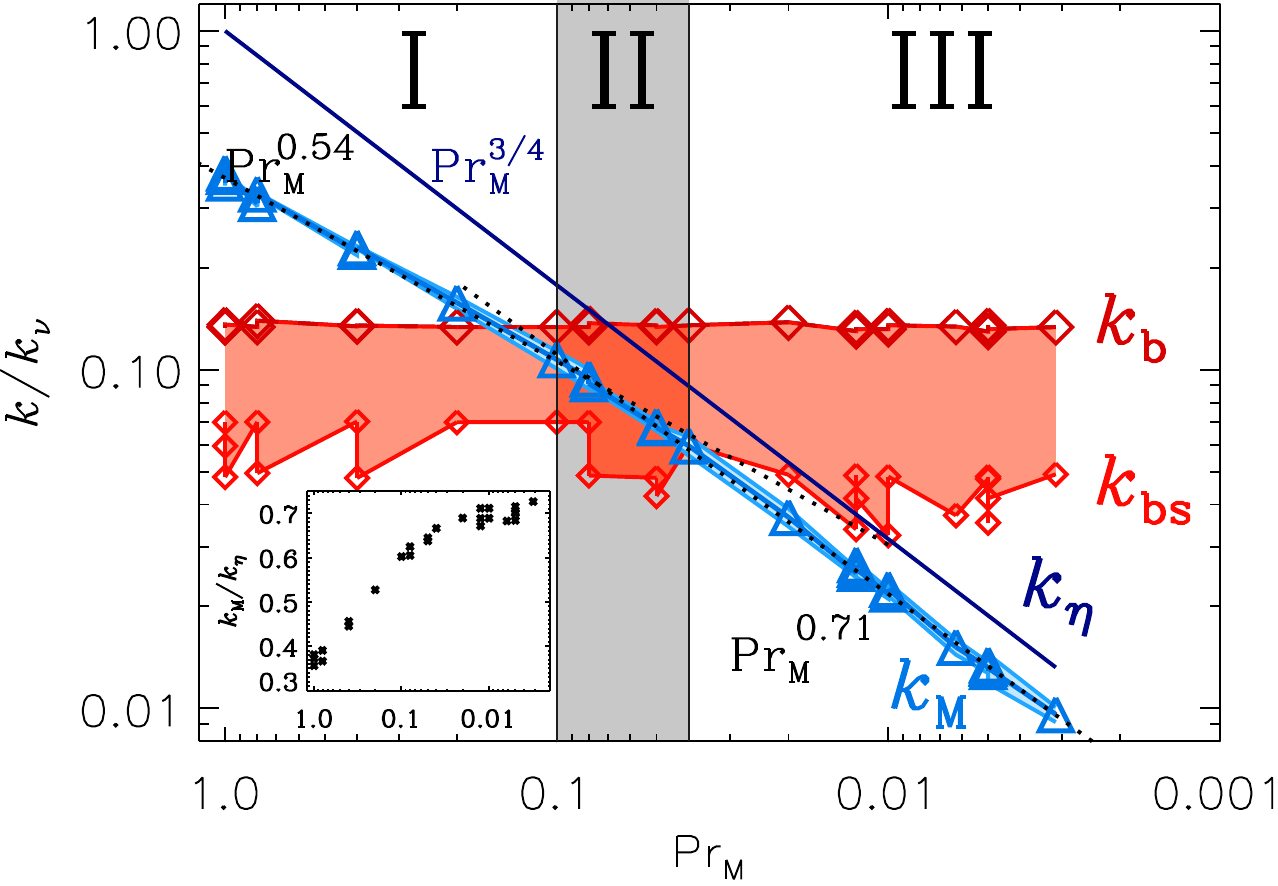}
\end{center}\caption[]{
{\bf Relation of the characteristic magnetic wavenumber $\kM$ to the bottleneck.}
We show its peak $\kB$ and its starting point $\kBs$ in red,
the characteristic magnetic wavenumber $\kM$ in light blue and the ohmic dissipation wavenumber $\ke$ in dark blue.
The red shaded area between $\kB$ and $\kBs$ corresponds to the low-wavenumber part of the bottleneck
where the turbulent flow is rougher than for a $-5/3$ power-law.
The Roman
numbers indicate the three distinct regions of dynamo excitation.
The region of the weakest
growth (II) is over-plotted in grey.
The characteristic magnetic wavenumber $\kM$ can be fitted by two power laws
(black dotted lines): $\kM/\kn\propto \Pm^{0.54}$ for $\Pm\ge0.05$  and $\kM/\kn\propto \Pm^{0.71}$ for
$\Pm\le0.05$.
All wavenumbers are normalized by the viscous one $\kn$.
We find that the dynamo is hardest to excite if $\kM$ lies within
the low-wavenumber side of the bottleneck.
Leaving this region towards lower or higher wavenumbers makes
the dynamo easier to excite.
The inset shows $\kM/\ke$ as a function of $\Pm$.
}\label{inter}
\end{figure}

\subsection*{Regions of dynamo excitation}
Next we seek answers to the obvious question arising: why is the SSD harder to excite in
a certain intermediate range of $\Pm$
and easier at lower and higher values?
For this, we investigate the
kinetic and magnetic energy spectra of a representative subset of the runs, 
see Supplementary Table~2.
We show in Fig.~\ref{spec} spectra of two exemplary cases:
Run F005, with $\Pm=0.05$, probes the $\Pm$ interval of impeded dynamo action, while Run H0005, with $\Pm=0.005$, is clearly outside it;
see Supplementary Fig.~1 and 2 for spectra of other cases.

In all cases
the kinetic energy clearly follows a Kolmogorov cascade with $E_{\rm kin}\propto
k^{-5/3}$ in the inertial range. 
When compensating with $k^{5/3}$, we find the well-known
bottleneck effect \cite{FG94,LM95}: 
a local increase in spectral energy, deviating from the power-law, as found
both in fluid experiments \cite{SJ93,SV94,KBB19} and
numerical studies \cite{DHYB03,DS10}.
It has been postulated to be detrimental to SSD growth  
\cite{BC04,BHLS18}.
For the magnetic spectrum on the other hand, 
yet clearly visible only for $\Pm\le0.005$,
we find a power-law
following $E_{\rm mag}\propto k^{-3}$.
A $3/2$ slope at low wavenumbers as predicted by \cite{K68} is seen only in the runs with $\Pm$
close to one, while for the intermediate and low $\Pm$ runs, the positive-slope
part of the spectrum shrinks to cover only the lowest $k$ values, and the steep
negative slopes at high $k$ values
become prominent.  
A steep negative slope
in the magnetic power spectra was also seen by
\cite{SOCMPY07} for $\Pm$ slightly below unity.
However, the authors propose a tentative
power of $-1$ given that the $-3$ slope is not yet clearly visible for their $\Pm$ values.

Analyzing our simulations, we adopt the following strategy:
For each spectrum,
we determine the wavenumber of the bottleneck, $\kB$, 
as the location of its maximum in
the (smoothed) compensated spectrum,
along with its starting point 
$\kBs<\kB$ at the location with  75\% 
of the maximum, see the middle-row panels
of Fig.~\ref{spec}.
We additionally calculate
a characteristic magnetic wavenumber, defined as 
$\kM = \int_k E_{\rm mag}(k) k\, dk/\int_k E_{\rm mag}(k)\, dk$,
which is often connected with the energy-carrying scale.
Furthermore, we calculate the viscous dissipation wavenumber $\kn=(\epsilon_{\rm
  K}/\nu^3)^{1/4}$ following Kolmogorov theory, where $\epsilon_{\rm K}$ is the viscous
dissipation rate
$2\nu \SSSS^2$ with the traceless rate-of-strain tensor of the flow, $\SSSS$.
From the relations between these four wavenumbers (listed in Supplementary Table~2),
we will draw insights about the observed
behavior of $\Rm^{\rm crit}$ with respect to $\Pm$.

We plot $\kB/\kn$ and $\kBs/\kn$ as functions 
of $\Pm$ in Fig.~\ref{inter}.
As is expected, $\kB/\kn$, 
or the ratio of the viscous scale to the scale of the bottleneck,
does not depend on $\Pm$, as 
the bottleneck is a purely hydrodynamic phenomenon.
The start of the bottleneck $\kBs$
should likewise
not depend on
$\Pm$, but the low $\Rey$ values for $\Pm=1$ to 0.1 lead 
to apparent thinner bottlenecks, hence an unsystematic weak dependency.
The red shaded area between $\kB$ and $\kBs$ is the low-wavenumber
part of bottleneck where the 
slope of the spectrum is larger (less negative) than $-5/3$ 
see Supplementary Table~2 for values 
of the modified slope $\alpha_{\rm b}$
and Supplementary, Section~1 for a discussion.
We note
that  $\alpha_{\rm b} \approx -1.3 \ldots -1.5$ and can thus deviate significantly from $-5/3$.
Overplotting the $\kM/\kn$ curve reveals
that it intersects with
the red shaded area exactly where the dynamo is hardest to excite
(region II).
This lets us conclude that the shallower slope
of the low-wavenumber part of
the bottleneck may indeed be responsible for enhancing $\Rm^{\rm crit}$ in the interval $0.04\le\Pm\le0.1$.
Using this plot, we can now clearly explain the
three regions of dynamo excitation.
For $0.1\le\Pm\le 1$ the low-wavenumber part of bottleneck and the characteristic magnetic
scale are completely decoupled.
This makes the SSD easy to excite (region I). 
For $0.04\le\Pm\le0.1$, 
(grey, region II), the dynamo is hardest to excite
because of the shallower slope of the kinetic spectra.
In region III, where $\Pm\le0.04$ the low-wavenumber part of bottleneck and the characteristic magnetic
scale are again completely decoupled making the dynamo easier to excite.

Further, we find that the dependence of $\kM/\kn$ on $\Pm$ 
also differs between the regions.
In region I $\kM/\kn$ depends on $\Pm$ via
$\kM/\kn\propto\Pm^{0.54}$ 
and in region II and III via
$\kM/\kn\propto\Pm^{0.71}$.
This becomes particularly interesting when comparing
the characteristic magnetic wavenumber
$\kM$ with the ohmic dissipation wavenumber
which is defined as $\ke=\kn\Pm^{3/4}$.
In region I, we find a significant difference of $\kM$ and $\ke$ in value and scaling.
However, in region III the scaling of $\kM$ comes very close to the $3/4$ scaling of $\ke$.
This behaviour can be even better seen in the inset of Fig.~\ref{inter}, where
the ratio $\kM/\ke$ is 0.3 for $\Pm=1$ and tends towards unity for decreasing $\Pm$, but is likely to
saturate below 0.75.

\section*{Discussion}

In conclusion, we find that the SSD is progressively easier to excite for 
magnetic Prandtl numbers below 0.04, in contrast to earlier findings,
and thus is very likely to exist 
in the Sun and other cool  stars.
Provided saturation at sufficiently high levels,
the SSD has been proposed to
strongly influence the
dynamics of solar-like stars: previous numerical studies, albeit at $\Pm \approx 1$, indicate that
this influence concerns for example the angular momentum transport \cite{KKOWB17,HK21},
and the LSD
\cite{TC13,BSB16,SB16,HRY16,VPKRSK21}.
Our kinematic study, however, only shows that a positive growth rate is possible
at very low $\Pm$, but not whether an SSD is able to generate
dynamically important field strengths.
As the $\Rm$ of the Sun and solar-like stars is several orders of magnitude higher than the
extrapolated $\Rmc$ value of 40, we yet expect dynamically important SSDs as indicated by $\Pm=1$ simulations \cite{HRY15s}. 
However, numerical simulations with $\Pm$ down to 0.01
show a decrease of the saturation strength with decreasing $\Pm$ \cite{B11}.

The results of our study are well in agreement with previous numerical studies
considering partly overlapping $\Pm$ ranges \cite{SCMM04,SOCMPY07,ISCM07, BHLS18}.
Those found some discrepancies with the Kazantsev theory \cite{K68} for low $\Pm$,
for example the narrowing down of the
positive Kazantsev spectrum at low and intermediate wavenumbers, and the emergence
of a negative slope instead at large wave numbers \cite{SOCMPY07}.
We could extend this regime to even lower $\Pm$
and therefore study these discrepancies further.
For $\Pm\le0.005$
we find that the magnetic spectrum shows a power-law scaling $k^{-3}$, which
is significantly steeper than the tentative 
 $k^{-1}$ one proposed in \cite{SOCMPY07} for 
$0.03\lesssim\Pm\lesssim0.07$ (but only for 8th-order hyperdiffusivity).
This latest
finding of such a steep power law in the magnetic spectrum
challenges the current theoretical predictions and might indicate that
the SSD operating at low $\Pm$ is fundamentally different from that at
$\Pm\approx1$.

Secondly, we find that
the growth rates near the onset follow
an $\ln(\Rm)$
dependence as predicted by \cite{KR97,KR12},
and not a $\Rm^{1/2}$ one
as would result from intertial-range-driven SSD \cite{BS05, SOCMPY07}.
We do not observe a tendency of the growth rate to become independent of $\Rm$ 
at the highest $\Pm$ either, which could
be an indication of an outer-scale driven SSD, as postulated by \cite{SOCMPY07}.
Furthermore, we find that the pre-factor of $\gamma\propto\ln(\Rm/\Rmc)$ is nearly
constant with its mean around 0.022, in agreement with
0.023 of \cite{BHLS18}.
A constant value means that the logarithmic scaling is independent of
$\Pm$ and seems to be of general validity.

Thirdly, we find that
the measured characteristic magnetic
wavenumber $\kM$ is always smaller than the estimated
$\ke$, 
and furthermore, $\kM$ is not always following the theory-predicted scaling of $\ke\propto\Pm^{3/4}$
with $\Pm$.
For the region I, where $\Pm$ is close to 1, 
this discrepancy is
up to a factor of three and
the deviation from the expected
$\Pm$-scaling is most significant here.
These discrepancies have been associated
with the numerical setups injecting energy at a forcing scale    
far larger than the dissipation scale, i.e. $\kf\ll\ke$
\cite{BS05}.
Furthermore, our runs in region I also have relatively 
low $\Rey$ and therefore numerical effects are not dismissible.
In region III (low $\Pm$),
$\kM/\ke$ is approaching the constant offset factor 0.75.
Hence, the scaling of $\kM/\kn$ with $\Pm$ gets close to the expected 
one.
This result again indicates that the SSD at low $\Pm$ is different from that at
$\Pm\approx1$.

An increase of  $\Rmc$ with decreasing $\Pm$
followed by an
asymptotic levelling-off for $\Pm\ll1$
was expected in the light of theory
and previous numerical studies.
Instead, we found non-monotonic behavior as function of $\Pm$; we could relate it 
to the hydrodynamical phenomenon of the bottleneck.
If the characteristic magnetic wavenumber lies in the 
positive-gradient part of the compensated spectrum, where the spectral slope is 
significantly
reduced from $-5/3$ to $\sim -1.4$,
the dynamo is hardest to excite
($0.1\ge\Pm\ge0.04$). For higher or lower $\Pm$, the dynamo
becomes increasingly easier to excite.
The local change in slope due to the bottleneck 
has often been related to an increase of
the ``roughness'' of the flow \cite{DHYB03,BS05, BHLS18},
which is expected to 
harden dynamo excitation based on
theoretical predictions \cite{BC04,Schober+12} 
from kinematic Kazantsev theory \cite{K68}.
In line with theory, the roughness-increasing part of the bottleneck
appears decisive in our results,  however, only when $\kM$ is used as a criterion.
The usage of $\ke$ would in contrast suggest that the peak of the
bottleneck is decisive \cite{BHLS18}.
Such interpretation appears incorrect, as the rough
estimate of $\ke$ employed here does not 
represent the magnetic spectrum
adequately and the peak of the bottleneck 
does not coincide with
the maximum of ``roughness''.

\section*{Online Methods}

\subsection*{Numerical setup}\label{sec:numset}
For our simulations, we use a cubic Cartesian box with edge length $L$
and solve the isothermal magnetohydrodynamic equations without gravity, similar as in \cite{B01,Hau04}.
\begin{eqnarray}
\frac{\DD \uu}{\DD t} &=& -\cs^2\nab\ln\rho + \JJ\times\BB/\rho +
                          \nab\cdot(2\rho\nu\SSSS)/\rho + \ff,\\
\frac{\partial \AAA}{\partial t} &=& \uu\times\BB + \eta\nab^2\AAA,\\
\frac{\DD \rho}{\DD t} &=& -\nab\cdot(\rho\uu),
\end{eqnarray}
where $\uu$ is the flow speed, $\cs$ is the sound speed, 
$\rho$ is the mass density, and
$\BB=\nab\times\AAA$
is the magnetic field with $\AAA$ being the vector potential.
$\JJ=\nab\times\BB/\mu_0$ is the current density with
 magnetic vacuum permeability $\mu_0$, while
$\nu$ and $\eta$ are constant kinematic viscosity and magnetic
diffusivity, respectively.
The rate-of-strain tensor
$\SSS_{ij}=(u_{i,j}+u_{j,i})/2 -\delta_{ij}\nab\cdot\uu/3$ is
traceless.
The forcing function $\ff$ provides
random white-in-time 
non-helical transversal plane waves, which are added in each time step to the
momentum equation, see \cite{Hau04} for details. The wavenumbers of the forcing
lie in a narrow band around $\kf=2k_1$ with $k_1=2\pi/L$.
Its amplitude is chosen such that the Mach number $\Ma=\urms/\cs$ is
always around $0.082$, where   
$\urms=\sqrt{\bra{\uu^2}_V}$ is the volume and time-averaged root-mean-square value.
The $\Ma$ values of all runs are listed in Supplementary Material Table~1.
To normalize the growth rate $\lambda$, we use an estimated turnover time
$\tau=1/\urms\kf\approx 6/k_1\cs
$.
The boundary conditions are periodic for all quantities and we
initialise the magnetic field with weak Gaussian noise.

Diffusion is controlled by the prescribed parameters $\nu$ and $\eta$. Accordingly,
 we define the fluid and magnetic Reynolds numbers with the forcing wavenumber $\kf$ as
\begin{equation}
\Rey={\urms}/{\nu\kf}, \quad \Rm={\urms}/{\eta\kf}.
\end{equation}
We performed numerical free decay experiments (see Supplementary Material, Section~7),
from which we confirm that the numerical diffusivities are negligible.

The spectral kinetic and magnetic energy densities are defined via
\begin{eqnarray}
\int_k E_{\rm kin}(k)\, dk &=& \urms^2\,\brac{\rho}_V/2,\\
\int_k E_{\rm mag}(k)\, dk &=& \Brms^2/2\mu_0,   
\end{eqnarray}
where $\Brms=\sqrt{\bra{\BB^2}_V}$ is the volume-averaged root-mean-square value
and $\brac{\rho}_V$ the volume-averaged density.

Our numerical setup employs
a drastically simplified model of turbulence compared to the actual one in the Sun.
There, turbulence is driven by stratified rotating convection 
being of course neither isothermal nor isotropic.
However, these simplifications 
were to date necessary when performing a parameter study at such high resolutions as we do.
Nevertheless, we can connect our study to solar parameters in
terms of $\Pm$ and $\Ma$.
Their chosen values best represent 
the weakly stratified layers within 
the bulk of the solar convection zone, where $\Pm\ll1$ and  $\Ma\ll1$. 
The anisotropy in the flow on small scales
is much weaker there than near the surface and
therefore close to our simplified setup.

\subsection*{Data availability}
Data for reproducing Figs. 2, 3, and 5 are included in the article and its
supplementary information files. 
The raw data (timeseries, spectra, slices, and snapshots) are provided through IDA/Fairdata service hosted at
CSC, Finland, under \url{https://doi.org/10.23729/206af669-07fd-4a30-9968-b4ded5003014}.
From the raw data, Figs. 1 and 4 can be reproduced.

\subsection*{Code availability}
We use the Pencil Code \cite{PC21} to perform all simulations, with
parallelized fast-Fourier-transforms to calculate the 
spectra on the fly \cite{Bou2020}. Pencil Code is freely available under
\url{https://github.com/pencil-code/}.

\subsection*{Acknowledgements} 
We acknowledge fruitful discussions with Axel Brandenburg, Igor Rogachevskii,  Alexander Schekochihin,
and Jennifer Schober during the Nordita program on ``Magnetic field evolution in low density or strongly stratified plasmas".
Computing resources from CSC during the Mahti pilot project 
and from Max Planck Computing and Data Facility (MPCDF)
are gratefully acknowledged.
This project including all authors,
has received funding from the European Research Council (ERC)
under the European Union's Horizon 2020 research and innovation
program (Project UniSDyn, grant agreement n:o 818665).
This work was done in collaboration with the COFFIES DRIVE Science Center.

\subsection*{Author Contributions Statement}
JW was leading, but all the authors contributed, to the design and
performing the numerical simulations. JW was leading the data
analysis. MJKL was in charge of acquiring the computational resources
from CSC. All the authors contributed to the interpretation of the
results and writing up the manuscript.

\subsection*{Competing Interests Statement}
The authors declare no competing interests.


\bibliography{paper.bib}




\section*{Supplementary material}

# Numerical evidence for a small-scale dynamo approaching solar magnetic Prandtl numbers

# Supplementary Material

Jörn Warnecke[1], Maarit J. Korpi-Lagg[2,1,3], Frederick A. Gent[2,4] and Matthias Rheinhardt[2]

[1] Max-Planck-Institut für Sonnensystemforschung, Justus-von-Liebig-Weg 3, D-37077 Göttingen, Germany
[2] Department of Computer Science, Aalto University, PO Box 15400, FI-00 076 Espoo, Finland
[3] Nordita, KTH Royal Institute of Technology & Stockholm University, Hannes Alfvéns väg 12, SE-11419, Sweden
[4] School of Mathematics, Statistics and Physics, Newcastle University, NE1 7RU, UK

March 13, 2023

## 1 Discussion on the roughness of the flow

We calculate the slope in the low-wavenumber part of the bottleneck by fitting $E_{\rm mag} \sim k^{\alpha_{\rm b}}$ in the interval $k_{\rm bs} \ldots k_{\rm b}$. As shown in Supplementary Fig. 1, the values of $\alpha_{\rm b}$ are significantly different from $-5/3$ without the bottleneck. Furthermore, we find no clear systematic dependence of $\alpha_{\rm b}$ on Re.

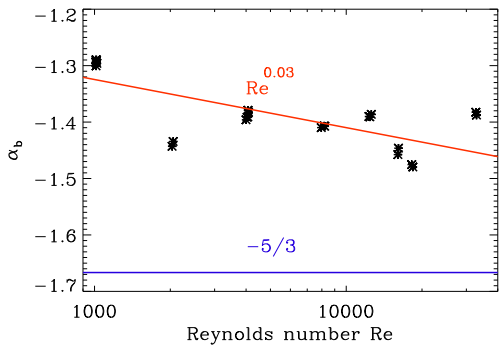

Supplementary Figure 1: Slope of the low-wavenumber part of the bottleneck, $\alpha_{\rm b}$, as a function of fluid Reynolds number Re. The blue line indicates the slope $-5/3$ in the inertial range without bottleneck; the red line shows a power-law fit with $\mathrm{Re}^{0.03}$.

## 2 Spectra of a representative subset of the runs

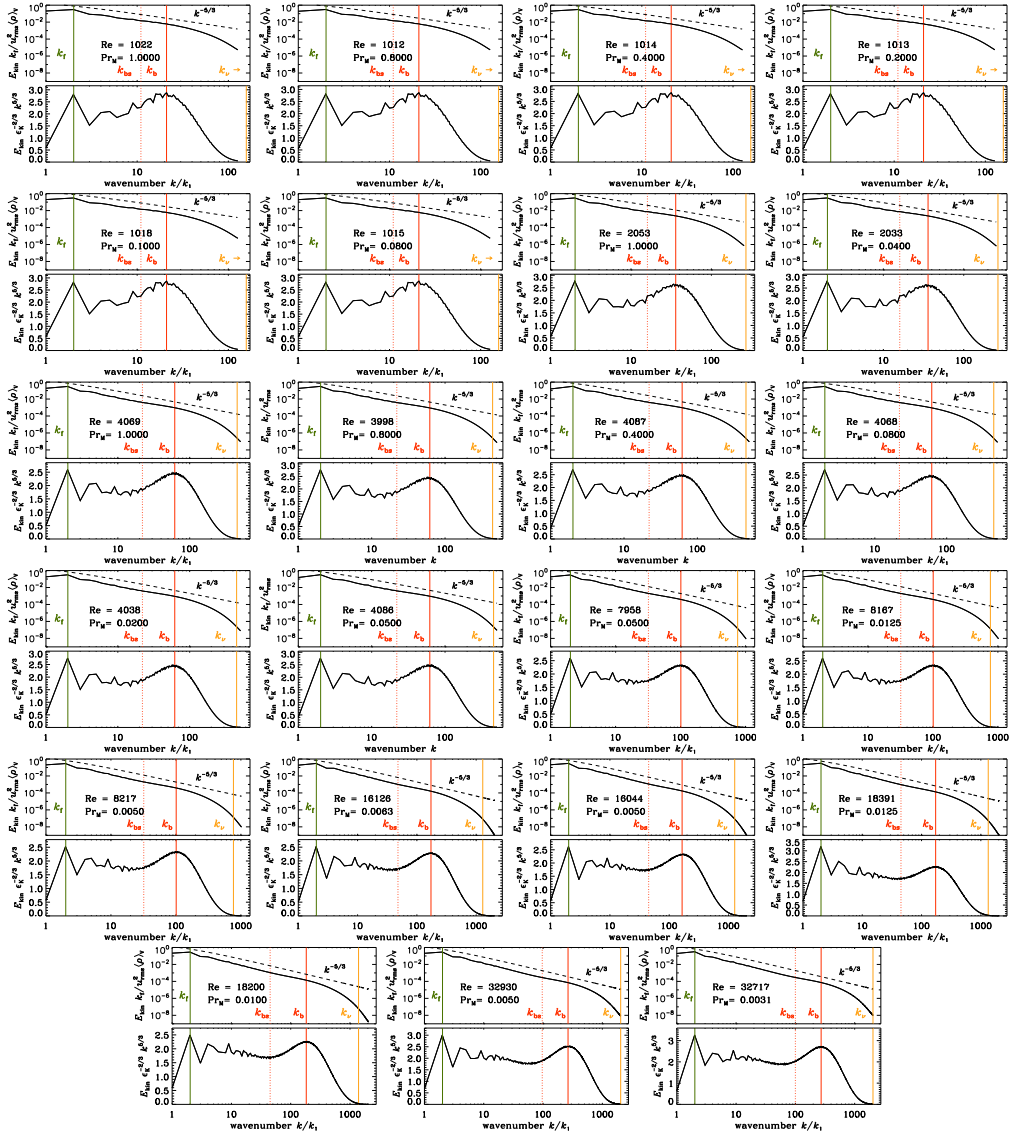

Supplementary Figure 2: Kinetic energy spectra, for explanations see Fig. 4. The notation "$k_\nu \to$" indicates that $k_\nu$ is outside the accessible $k$ range.

2

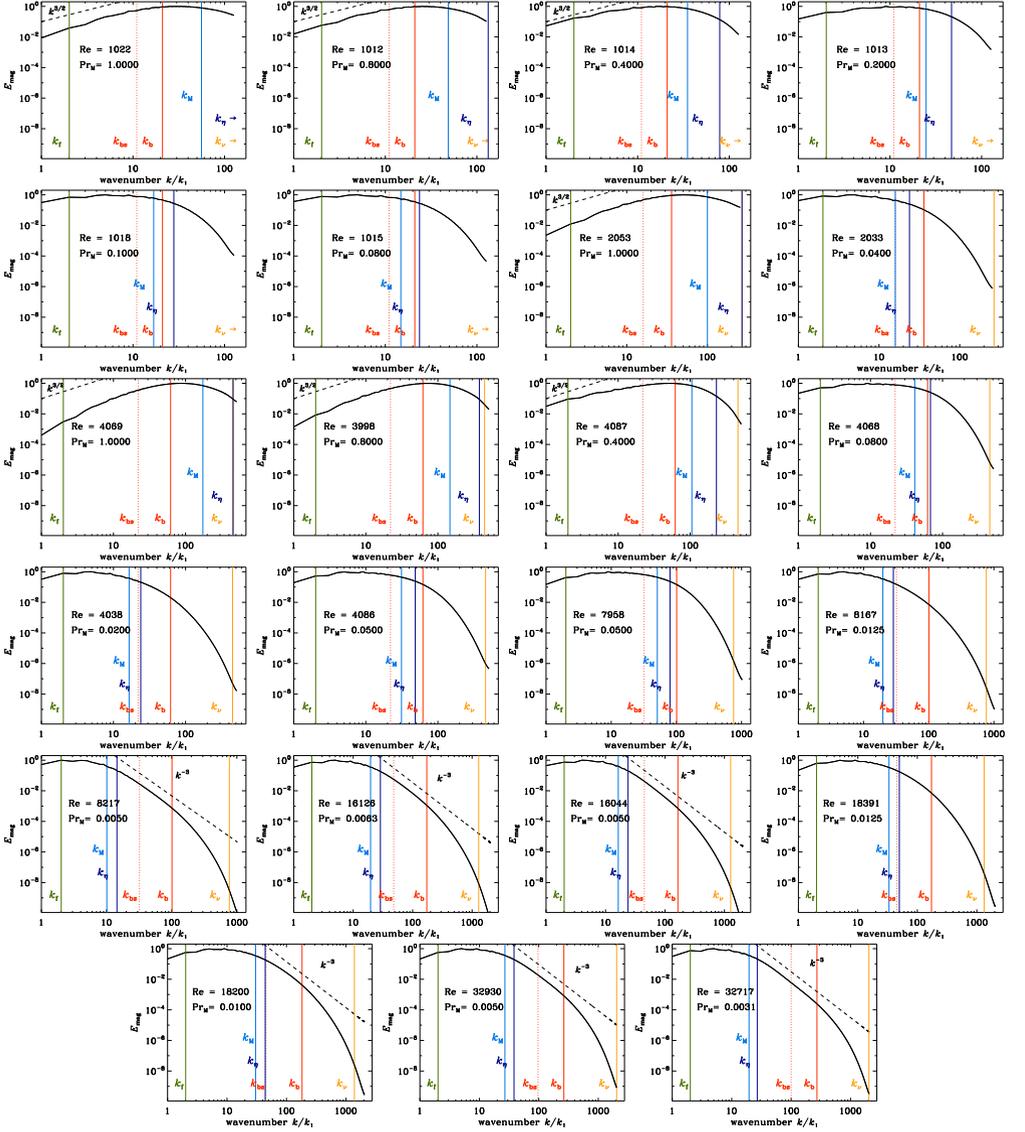

Supplementary Figure 3: Normalized magnetic energy spectra, for explanations see Fig 4. The notation "$k_{\nu,\eta} \to$" indicates that $k_{\nu,\eta}$ is outside the accessible $k$ range.

3

# 3 Table of all runs

Supplementary Table 1: Overview of all performed runs.

| Run | Resolution | $Pr_M$ | Re | $Re_M$ | Ma | $\lambda\tau$ | $\sigma_\lambda\tau$ |
|---|---|---|---|---|---|---|---|
| P10 | $256^3$ | 1.00 | 46 | 46 | 0.075 | 0.0102 | 0.0015 |
| A80 | $256^3$ | 8.00 | 99 | 790 | 0.080 | 0.0962 | 0.0013 |
| A40 | $256^3$ | 4.00 | 99 | 394 | 0.080 | 0.0794 | 0.0016 |
| A20 | $256^3$ | 2.00 | 99 | 197 | 0.080 | 0.0557 | 0.0009 |
| A15 | $256^3$ | 1.50 | 99 | 148 | 0.080 | 0.0458 | 0.0010 |
| A10 | $256^3$ | 1.00 | 99 | 99 | 0.080 | 0.0315 | 0.0007 |
| A08 | $256^3$ | 0.80 | 98 | 78 | 0.079 | 0.0200 | 0.0008 |
| A05 | $256^3$ | 0.50 | 98 | 49 | 0.079 | -0.0008 | 0.0019 |
| A04 | $256^3$ | 0.40 | 97 | 39 | 0.079 | -0.0052 | 0.0016 |
| B20 | $256^3$ | 2.00 | 405 | 811 | 0.081 | 0.1129 | 0.0010 |
| B10 | $256^3$ | 1.00 | 404 | 404 | 0.081 | 0.0707 | 0.0004 |
| B05 | $256^3$ | 0.50 | 404 | 202 | 0.081 | 0.0334 | 0.0019 |
| B0375 | $256^3$ | 0.375 | 403 | 151 | 0.081 | 0.0182 | 0.0011 |
| B025 | $256^3$ | 0.25 | 405 | 101 | 0.081 | 0.0015 | 0.0010 |
| B02 | $256^3$ | 0.2 | 405 | 81 | 0.082 | -0.0060 | 0.0025 |
| B0125 | $256^3$ | 0.125 | 405 | 51 | 0.082 | -0.0217 | 0.0025 |
| B01 | $256^3$ | 0.1 | 404 | 40 | 0.082 | -0.0287 | 0.0015 |
| C10 | $256^3$ | 1.00 | 1022 | 1022 | 0.082 | 0.1428 | 0.0075 |
| C08 | $256^3$ | 0.80 | 1014 | 811 | 0.082 | 0.1079 | 0.0028 |
| C04 | $256^3$ | 0.40 | 1013 | 405 | 0.082 | 0.0486 | 0.0007 |
| C02 | $256^3$ | 0.20 | 1008 | 201 | 0.081 | 0.0132 | 0.0009 |
| C015 | $256^3$ | 0.15 | 1014 | 152 | 0.082 | 0.0027 | 0.0008 |
| C01 | $256^3$ | 0.10 | 1013 | 101 | 0.081 | -0.0074 | 0.0010 |
| C008 | $256^3$ | 0.08 | 1015 | 81 | 0.082 | -0.0154 | 0.0022 |
| C005 | $256^3$ | 0.05 | 1019 | 51 | 0.082 | -0.0239 | 0.0014 |
| D10 | $512^3$ | 1.00 | 2053 | 2053 | 0.083 | 0.1853 | 0.0052 |
| D04 | $512^3$ | 0.40 | 2038 | 815 | 0.082 | 0.0682 | 0.0017 |
| D02 | $512^3$ | 0.20 | 2026 | 405 | 0.082 | 0.0235 | 0.0006 |
| D01 | $512^3$ | 0.10 | 2038 | 204 | 0.082 | 0.0053 | 0.0021 |
| D0075 | $512^3$ | 0.075 | 2029 | 152 | 0.082 | -0.0005 | 0.0015 |
| D005 | $512^3$ | 0.05 | 2031 | 102 | 0.082 | -0.0058 | 0.0017 |
| D004 | $512^3$ | 0.04 | 2034 | 81 | 0.082 | -0.0121 | 0.0009 |
| D0025 | $512^3$ | 0.025 | 2030 | 51 | 0.082 | -0.0236 | 0.0017 |
| E10 | $1024^3$ | 1.00 | 4069 | 4069 | 0.082 | 0.2769 | 0.0126 |
| E08 | $1024^3$ | 0.80 | 3999 | 3199 | 0.081 | 0.2141 | 0.0091 |
| E04 | $1024^3$ | 0.40 | 4087 | 1635 | 0.082 | 0.1148 | 0.0324 |
| E02 | $1024^3$ | 0.20 | 4064 | 813 | 0.082 | 0.0401 | 0.0025 |
| E01 | $1024^3$ | 0.10 | 4075 | 407 | 0.082 | 0.0155 | 0.0009 |
| E008 | $1024^3$ | 0.08 | 4089 | 325 | 0.082 | 0.0191 | 0.0025 |
| E005 | $1024^3$ | 0.05 | 4090 | 205 | 0.082 | 0.0044 | 0.0021 |
| E00375 | $1024^3$ | 0.0375 | 4121 | 155 | 0.083 | 0.0006 | 0.0024 |
| E003 | $1024^3$ | 0.03 | 4084 | 123 | 0.082 | -0.0035 | 0.0012 |
| E002 | $1024^3$ | 0.02 | 4037 | 81 | 0.081 | -0.0163 | 0.0023 |
| E001 | $1024^3$ | 0.01 | 4043 | 40 | 0.082 | -0.0232 | 0.0029 |

$Pr_M$ is the magnetic Prandtl number, Re and $Re_M$ are the fluid and magnetic Reynolds numbers, Ma = $u_{rms}/c_s$ is the Mach number, $\tau = 1/u_{rms}k_f$ is a rough estimate for the turnover time, $\lambda$ is the SSD growth rate with its error $\sigma_\lambda$.

Supplementary Table 1: continued.

| Run | Resolution | Pr$_M$ | Re | Re$_M$ | Ma | $\lambda\tau$ | $\sigma_\lambda\tau$ |
|---|---|---|---|---|---|---|---|
| F01 | $2048^3$ | 0.10 | 8028 | 803 | 0.081 | 0.0320 | 0.0046 |
| F005 | $2048^3$ | 0.05 | 7959 | 398 | 0.080 | 0.0192 | 0.0029 |
| F0025 | $2048^3$ | 0.025 | 8050 | 201 | 0.081 | 0.0108 | 0.0026 |
| F001875 | $2048^3$ | 0.01875 | 8093 | 152 | 0.082 | 0.0039 | 0.0027 |
| F00125 | $2048^3$ | 0.0125 | 8171 | 102 | 0.082 | -0.0004 | 0.0022 |
| F001 | $2048^3$ | 0.01 | 8178 | 82 | 0.082 | -0.0009 | 0.0033 |
| F0005 | $2048^3$ | 0.005 | 8217 | 41 | 0.083 | -0.0162 | 0.0035 |
| FG00125 | $2048^3$ | 0.0125 | 12580 | 157 | 0.084 | 0.0026 | 0.0015 |
| FG001 | $2048^3$ | 0.01 | 12313 | 123 | 0.082 | -0.0041 | 0.0021 |
| FG0005 | $2048^3$ | 0.005 | 12415 | 62 | 0.083 | -0.0146 | 0.0039 |
| G00125 | $4096^3$ | 0.0125 | 18391 | 229 | 0.093 | 0.0372 | 0.0063 |
| G001 | $4096^3$ | 0.01 | 18200 | 182 | 0.092 | 0.0065 | 0.0062 |
| G000625 | $4096^3$ | 0.00625 | 16126 | 101 | 0.081 | -0.0049 | 0.0069 |
| G0005 | $4096^3$ | 0.005 | 16071 | 80 | 0.081 | -0.0152 | 0.0008 |
| H0005 | $4096^3$ | 0.005 | 32930 | 165 | 0.083 | 0.0103 | 0.0019 |
| H00031 | $4096^3$ | 0.003125 | 32717 | 102 | 0.082 | 0.0021 | 0.0052 |
| H00025 | $4608^3$ | 0.0025 | 32910 | 82 | 0.083 | -0.0097 | 0.0035 |

# 4 Table of spectral properties of a subset of the runs

Supplementary Table 2: Selected runs with spectral properties.

| $Pr_M$ | Re | $Re_M$ | $k_\nu$ | $k_{bs}$ | $k_b$ | $k_M$ | $k_\eta$ | $\alpha_b$ |
|---|---|---|---|---|---|---|---|---|
| 1.000000 | 1022 | 1022 | 157 | 11 | 21 | 56 | 157 | -1.30 |
| 1.000000 | 2053 | 2053 | 268 | 16 | 36 | 100 | 268 | -1.43 |
| 1.000000 | 4069 | 4069 | 456 | 22 | 62 | 173 | 456 | -1.38 |
| 0.800000 | 1012 | 810 | 157 | 11 | 21 | 48 | 132 | -1.29 |
| 0.800000 | 3998 | 3198 | 444 | 22 | 62 | 146 | 376 | -1.40 |
| 0.400000 | 1014 | 405 | 156 | 11 | 21 | 35 | 78 | -1.30 |
| 0.400000 | 4087 | 1634 | 459 | 22 | 62 | 105 | 231 | -1.38 |
| 0.200000 | 1013 | 202 | 157 | 11 | 21 | 24 | 47 | -1.29 |
| 0.100000 | 1018 | 101 | 157 | 11 | 21 | 16 | 27 | -1.29 |
| 0.080000 | 1015 | 81 | 157 | 11 | 21 | 14 | 23 | -1.29 |
| 0.080000 | 4068 | 325 | 451 | 22 | 62 | 41 | 67 | -1.39 |
| 0.050000 | 4086 | 204 | 458 | 22 | 62 | 31 | 48 | -1.38 |
| 0.050000 | 7958 | 397 | 755 | 32 | 101 | 50 | 79 | -1.41 |
| 0.040000 | 2033 | 81 | 266 | 16 | 36 | 15 | 23 | -1.44 |
| 0.020000 | 4038 | 80 | 449 | 22 | 62 | 16 | 23 | -1.39 |
| 0.012500 | 8167 | 102 | 766 | 32 | 101 | 19 | 28 | -1.41 |
| 0.012500 | 12580 | 157 | 1026 | 50 | 137 | 27 | 38 | -1.39 |
| 0.012500 | 18391 | 229 | 1332 | 45 | 173 | 33 | 49 | -1.48 |
| 0.010000 | 12313 | 123 | 1013 | 49 | 137 | 22 | 32 | -1.39 |
| 0.010000 | 18200 | 182 | 1388 | 45 | 183 | 30 | 43 | -1.48 |
| 0.006250 | 16126 | 100 | 1291 | 48 | 173 | 19 | 28 | -1.45 |
| 0.005000 | 8217 | 41 | 767 | 32 | 101 | 10 | 14 | -1.41 |
| 0.005000 | 12415 | 62 | 1013 | 49 | 137 | 13 | 19 | -1.39 |
| 0.005000 | 16044 | 80 | 1273 | 45 | 165 | 16 | 23 | -1.46 |
| 0.005000 | 32930 | 164 | 2037 | 97 | 264 | 26 | 38 | -1.39 |
| 0.003125 | 32717 | 102 | 2036 | 100 | 272 | 19 | 26 | -1.38 |

$Pr_M$ is the magnetic Prandtl number, Re and $Re_M$ are the fluid and magnetic Reynolds numbers, respectively. $k_\nu = (\epsilon_K/\nu^3)^{1/4}$ is the viscous dissipation wavenumber with the viscous dissipation rate $\epsilon_K$. $k_b$ and $k_{bs}$ locate the maximum of the bottleneck and its starting point (75% of the maximum), respectively. $k_M = \int_k E_{mag}(k)k\,dk / \int_k E_{mag}(k)dk$ is the characteristic magnetic wavenumber. $k_\eta = k_\nu Pr_M^{3/4}$ is the ohmic dissipation wavenumber and $\alpha_b$ is the slope of the low-wavenumber part of the bottleneck. All wavenumbers in units of $k_1$.

# 5 Details of fitting procedures and error calculation

## 5.1 Growth rates

The growth rate $\lambda$ is calculated from the time series of $B_\text{rms}$ of the hydrodynamically saturated stage. For this, we fit $B_\text{rms}(t)$ with $\exp(\lambda t)$ using a least-$\chi^2$ fit. For a small number of runs, mostly at $\text{Pr}_\text{M} \approx 1$, we need to exclude that part of the time series, where no longer $B_\text{rms} \ll \sqrt{\rho u_\text{rms}^2}$ holds, i.e. the kinematic stage is left. The majority of runs always stays in the kinematic stage though. The error of the growth rate, $\sigma_\lambda$, is estimated from the maximum deviation of the fit, which is determined as follows: the time series is divided into three equal parts, each fitted to obtain its value $\lambda_i$, $i = 1, \ldots, 3$. We then use half the maximum difference to $\lambda$ for the error, i.e. $\sigma_\lambda = \max |\lambda_i - \lambda|/2$. The minimal length of the time series was chosen to guarantee that $\sigma_\lambda$ is small enough while the maximum is bound by the resolution, hence the computational cost. For resolutions of $256^3 - 512^3$, we have $t_\text{max}/\tau \approx 500$, for $1024^3 - 2048^3$, $t_\text{max}/\tau \approx 150$ and for $4096^3$, $t_\text{max}/\tau \approx 15$. $\lambda$ and its error $\sigma_\lambda$ are listed in Supplementary Table 1.

## 5.2 Critical magnetic Reynolds number

To calculate the critical magnetic Reynolds number $\text{Re}_\text{M}^\text{crit}$ for each $\text{Pr}_\text{M}$ set, we fit a logarithmic function to the growth rate, i.e. $\lambda\tau = C_1 \ln(\text{Re}_\text{M}) + C_2$. This functional form is motivated by the theoretical expectation of [4, 1] and also by the distribution of the growth rates in Fig. 3a. We again use a least-$\chi^2$ fit, but take into account the growth rate errors, $\sigma_\lambda$, for the weights. It turns out that the pre-factor $C_1$ varies between 0.017 and 0.028 with a mean of 0.022. Thus, we conclude that the chosen functional form is appropriate. $\text{Re}_\text{M}^\text{crit}(\text{Pr}_\text{M})$ is calculated analytically from the zero of the fitting function, $\text{Re}_\text{M}^\text{crit} = \exp(-C_2/C_1)$, and its error $\sigma_{\text{Re}_\text{M}^\text{crit}}$ is derived directly from each fit, representing a one-sigma uncertainty of $\text{Re}_\text{M}^\text{crit}$. The values and errors of $\text{Re}_\text{M}^\text{crit}$ are then used in Fig. 3b and are listed in Supplementary Table 3 together with $C_1$. For the relative error $\sigma_{\text{Re}_\text{M}^\text{crit}}/\text{Re}_\text{M}^\text{crit}$, one finds for most of the runs values around $10 - 20\%$. However, taking these errors into account the decrease of $\text{Re}_\text{M}^\text{crit}$ as function of $\text{Pr}_\text{M}$ is still significant.

Supplementary Table 3: Critical magnetic Reynolds number

| $\text{Pr}_\text{M}$ | $\text{Re}_\text{M}^\text{crit}$ | $\sigma_{\text{Re}_\text{M}^\text{crit}}$ | $C_1$ |
|---|---|---|---|
| 1.0000 | 32 | 1.0 | 0.028 |
| 0.4000 | 60 | 2.6 | 0.025 |
| 0.2000 | 101 | 10.1 | 0.017 |
| 0.1000 | 150 | 11.7 | 0.021 |
| 0.0800 | 153 | 21.1 | 0.025 |
| 0.0500 | 153 | 13.4 | 0.021 |
| 0.0250 | 138 | 15.3 | 0.024 |
| 0.0125 | 119 | 49.8 | 0.018 |
| 0.0100 | 136 | 31.8 | 0.017 |
| 0.0050 | 108 | 18.5 | 0.022 |
| 0.003125 | 102 | - | - |

$\text{Re}_\text{M}^\text{crit}$ is the critical magnetic magnetic Reynolds number obtained by a fit $\lambda\tau = C_1 \ln(\text{Re}_\text{M}/\text{Re}_\text{M}^\text{crit})$ to the growth rate and $\sigma_{\text{Re}_\text{M}^\text{crit}}$ its error. The last row refers to the only run for $\text{Pr}_\text{M} = 0.003125$ with $\lambda \approx 0$, hence $\text{Re}_\text{M} \approx \text{Re}_\text{M}^\text{crit}$ (no fit).

## 5.3 Other fits

To obtain $\text{Re}_\text{M}^\text{crit}(\text{Pr}_\text{M}) \propto \text{Pr}_\text{M}^{0.125}$ in Fig. 3b, we use also a least-$\chi^2$ fit, employing the errors $\sigma_{\text{Re}_\text{M}^\text{crit}}$, similar for $k_\text{M}/k_\nu \propto \text{Pr}_\text{M}^\alpha$ in Fig 4.

# 6 Magnetic energy transfer functions

Seeking further insight into the dynamo operating in the three regimes of $Pr_M$, we look at the spectral magnetic energy transfer functions. We follow the approach of [2], but using the convention by [3], where the contribution of compressibility is subsumed in the stretching/shearing and advection terms, $T_{Str}$ and $T_{Adv}$, respectively, as follows

$$T_{Str}(\boldsymbol{k}) = \hat{\boldsymbol{B}}(\boldsymbol{k}) \cdot \widehat{(+\boldsymbol{B}\cdot\boldsymbol{\nabla u} - \boldsymbol{B}\boldsymbol{\nabla}\cdot\boldsymbol{u}/2)}^{*}(\boldsymbol{k})/2\mu_0 + \text{c.c.} \tag{1}$$

$$T_{Adv}(\boldsymbol{k}) = \hat{\boldsymbol{B}}(\boldsymbol{k}) \cdot \widehat{(-\boldsymbol{u}\cdot\boldsymbol{\nabla B} - \boldsymbol{B}\boldsymbol{\nabla}\cdot\boldsymbol{u}/2)}^{*}(\boldsymbol{k})/2\mu_0 + \text{c.c.} \tag{2}$$

Here, the hats indicate the Fourier transform and c.c. refers to the complex conjugate expressions. In Supplementary Fig. 4a, we show these functions after shell integration, i.e. as functions of $k = |\boldsymbol{k}|$, for six runs with $0.005 \leq Pr_M \leq 0.8$. As expected, the curves peak at higher wavenumbers for higher $Pr_M$. We also look at the ratio of the wavenumbers at which $T_{Str}$ and $T_{Adv}$ has its maximum and minimum, respectively, $k_{Str}/k_{Adv}$, which is determined from the curves after smoothing. We find that this ratio is maximal where the dynamo is hardest to excite, shown as grey shade in the Supplementary Fig. 4b. However, we should note that the simulations with the two lowest $Pr_M$ have a higher resolution than the other four, which could also influence the result. It remains unclear whether, or how, this enhanced ratio would relate to the difficulty in exciting the dynamo at $0.1 \geq Pr_M \geq 0.04$.

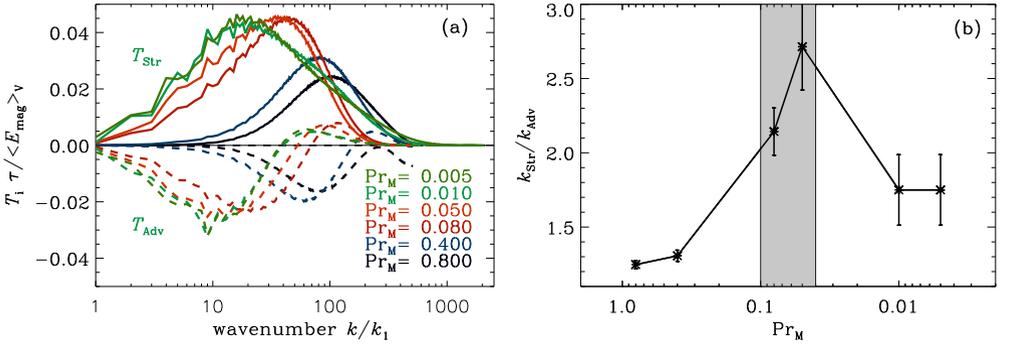

Supplementary Figure 4: Shell integrated spectral magnetic energy transfer functions vs. magnetic Prandtl number $Pr_M$. Panel a: transfer by stretching/shearing, $T_{Str}$, and by advection, $T_{Adv}$, normalized by the total magnetic energy density $\langle E_{mag}\rangle_V = \int_k E_{mag}(k)\,dk$ divided by turnover time $\tau$ and averaged over time. Color coding refers to $Pr_M$ as shown in the legend. Panel b: ratio of the wavenumbers of maximal stretching and minimal advection, $k_{Str}/k_{Adv}$, as a function of $Pr_M$. The grey shaded area indicates the interval $0.1 \geq Pr_M \geq 0.04$, where the dynamo is hardest to excite. The errors are calculated using an estimated uncertainty in determining $k_{Str}$ and $k_{Adv}$. Used runs: E08, E04, E008, E005, G001, H0005.

8

# 7 Numerical diffusivity

In our low $\mathrm{Pr_M}$ simulations, $\eta \gg \nu$, hence numerical diffusion might only play a role for the velocity field. We estimate the numerical viscosity in our simulations in two ways. Firstly, we follow the approach of [2], to estimate the fluid Reynolds number from the power spectra. For this we determine the Taylor microscale

$$\lambda_{\mathrm{TM}} = \left( \int_k E_{\mathrm{kin}}(k) dk \Big/ \int_k k^2 E_{\mathrm{kin}}(k) dk \right)^{1/2}, \qquad (3)$$

and the integral scale for turbulent motions

$$\lambda_{\mathrm{int}} = \int_k k^{-1} E_{\mathrm{kin}}(k) dk \Big/ \int_k E_{\mathrm{kin}}(k) dk. \qquad (4)$$

These scales can be used to estimate the effective Reynolds number as

$$\mathrm{Re}_{\mathrm{eff}} \propto (\lambda_{\mathrm{int}}/\lambda_{\mathrm{TM}})^2. \qquad (5)$$

This is often used to estimate $\mathrm{Re}_{\mathrm{eff}}$ from observations or in simulations that use only numerical diffusivities, although the proportionality constant is unknown. As shown in Supplementary Fig. 5, the effective Re based on Equation (5) scales linearly with Re based on the prescribed value of $\nu$. Assuming now plausibly numerical diffusion to enter mainly in the form of hyperdiffusion, we infer that from this linear relationship a notable influence of numerical diffusion can be ruled out.

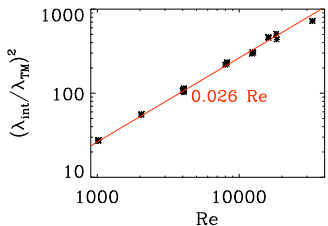

Supplementary Figure 5: Effective Reynolds number proxy (5) as a function of Re from the prescribed $\nu$. The red line indicates $(\lambda_{\mathrm{int}}/\lambda_{\mathrm{TM}})^2 = 0.026\,\mathrm{Re}$.

In addition, we performed a flow decay experiment to estimate the numerical diffusion from the flow decay rate in the absence of all other terms. The resulting effective viscosity is to high precision the same as the prescribed $\nu$ up to wavenumbers $k \lesssim 1000\,k_1$, covering safely all scales relevant for the magnetic field. From these two results we conclude that our numerical simulations are not affected by numerical diffusion, and that the $\mathrm{Pr_M}$ regimes are accurately identified.



\end{document}